\documentclass[prl,twocolumn,superscriptaddress,amsmath,amssymb,notitlepage]{revtex4}
\usepackage{graphicx}
\usepackage{amsmath}
\usepackage{amssymb}
\usepackage{float}
\usepackage{color}

\usepackage{cancel,xcolor}
\usepackage{indentfirst}
\usepackage{CJKulem}
\usepackage{soul,xcolor}
\setstcolor{red}

\usepackage{ulem}
\usepackage{cancel}
\usepackage[pdfpagemode=UseNone,pdfstartview=FitH,colorlinks=true,linkcolor=blue,urlcolor=blue,anchorcolor=blue,citecolor=blue]{hyperref}

\newcommand{\beq}{\begin{equation}}
\newcommand{\eeq}{\end{equation}}
\newcommand{\beqa}{\begin{eqnarray}}
\newcommand{\eeqa}{\end{eqnarray}}

\newcommand{\MUL}{K. A. M\"uller}

\def\Aa2#1{\textcolor{magenta}{#1}}

\def\Aa1#1{\textcolor{blue}{#1}}
\def\prb#1{{ Phys.\ Rev. B\/} {\bf#1}}

\def\prl#1{{ Phys.\ Rev.\ Lett.} {\bf#1}}

\begin{document}

\title{My encounters with Alex M\"{u}ller and the perovskites}

\author{Amnon Aharony}
\email{aaharonyaa@gmail.com}
\affiliation{ School of Physics and Astronomy, Tel Aviv University, Tel Aviv 6997801, Israel}

\begin{abstract}

This paper is dedicated to the memory of Professor K. Alex M\"uller. After describing our personal and scientific encounters since 1974, I concentrate on the many puzzles which appeared in our discussions and collaborations,  involving the  interplay between theory and experiments on the critical behavior of cubic perovskites which undergo (second or first) order transitions to a lower symmetry phases (trigonal or tetragonal). The conclusion, reached only very recently, is that (although beginning with the same cubic symmetry) the two types of transitions belong to two distinct universality classes: under [100] stress, the cubic to trigonal transition exhibits a tetracritical phase diagram, with `cubic' exponents, while the cubic to tetragonal transition exhibit an `intermediate' bicritical phase diagram, but asymptotically the bicritical point turns into a triple point, with three first order lines. To test these conclusions, it is suggested to measure the effective critical exponents as the temperature approaches criticality.

\end{abstract}

%%%%%%%%%%%%%%%%%%%%%%%%%%%%%%%%%%%%%%%%%%%%%%%%%
%%%%%%%%%%%%%%%%%%%%%%%%%%%%%%%%%%%%%%%%%%%%%%%%%

\date{\today}
\maketitle
%%%%%%%%%%%%%%%%%%%%%%%%%%%%%%%%%%%%%%%%%%%%%%%%%
%%%%%%%%%%%%%%%%%%%%%%%%%%%%%%%%%%%%%%%%%%%%%%%%%

\section{}
\vspace{-0.9 cm}
\subsection{I. Personal}

My discussions and collaborations with Karl Alexander M\"uller (Alex), which started in 1974 and continued into the 1990's, all concerned the critical phenomena near the displacive phase transitions of the perovskites (from a cubic to lower-symmetry structures). This paper concentrates on these encounters. Alex started his work on the perovskites in his Ph. D. thesis, published in 1958~\cite{phd}. In this thesis he measured EPR (electron paramagnetic resonance) spectra of Fe$^{3+}$ impurities in the perovskite SrTiO$^{}_3$, a tool which he continued to use for many important later discoveries, concerning these critical phenomena. Due to his work on the perovskites, Alex quickly became one of the leaders in the fields of structural transitions and ferroelectricity, and participated in many  international conferences and schools on these topics~\cite{NATO}. Much of the work on these topics is described in the 1991 two volume collection of papers, which he edited together with Harry Thomas~\cite{KAMrev}. Fifty of Alex's papers were reprinted in Ref. \cite{KAMbook}. Based on these contributions, Alex became an IBM Fellow in 1982. This gave him the right to choose his own research, eventually discovering high temperature superconductivity.

   Personally, I first met Alex when he visited Cornell, in 1974. At the time, I was a postdoc in Michael Fisher's group, working on applications of the (then) new renormalization group (RG), and I had just published a paper on the critical behavior of cubic systems~\cite{AA1973}. During his visit, Alex described his experimental results~\cite{KAM71}, and we  immediately started discussions on the interplay between the theory for the cubic perovskites and the experiments, which raised several puzzles. Many of these discussions were joint with Alastair Bruce (ADB), a former student of Roger Cowley in Edinburgh (who was also an expert on structural phase transitions) who was also a postdoc at Cornell. This `ping pong' exchange between Alex and me continued for more than 20 years (into his superconductivity years), during which Alex and I met many times, and became good friends~\cite{paris}.

\begin{figure}
\centering
\includegraphics[width=0.4\textwidth]{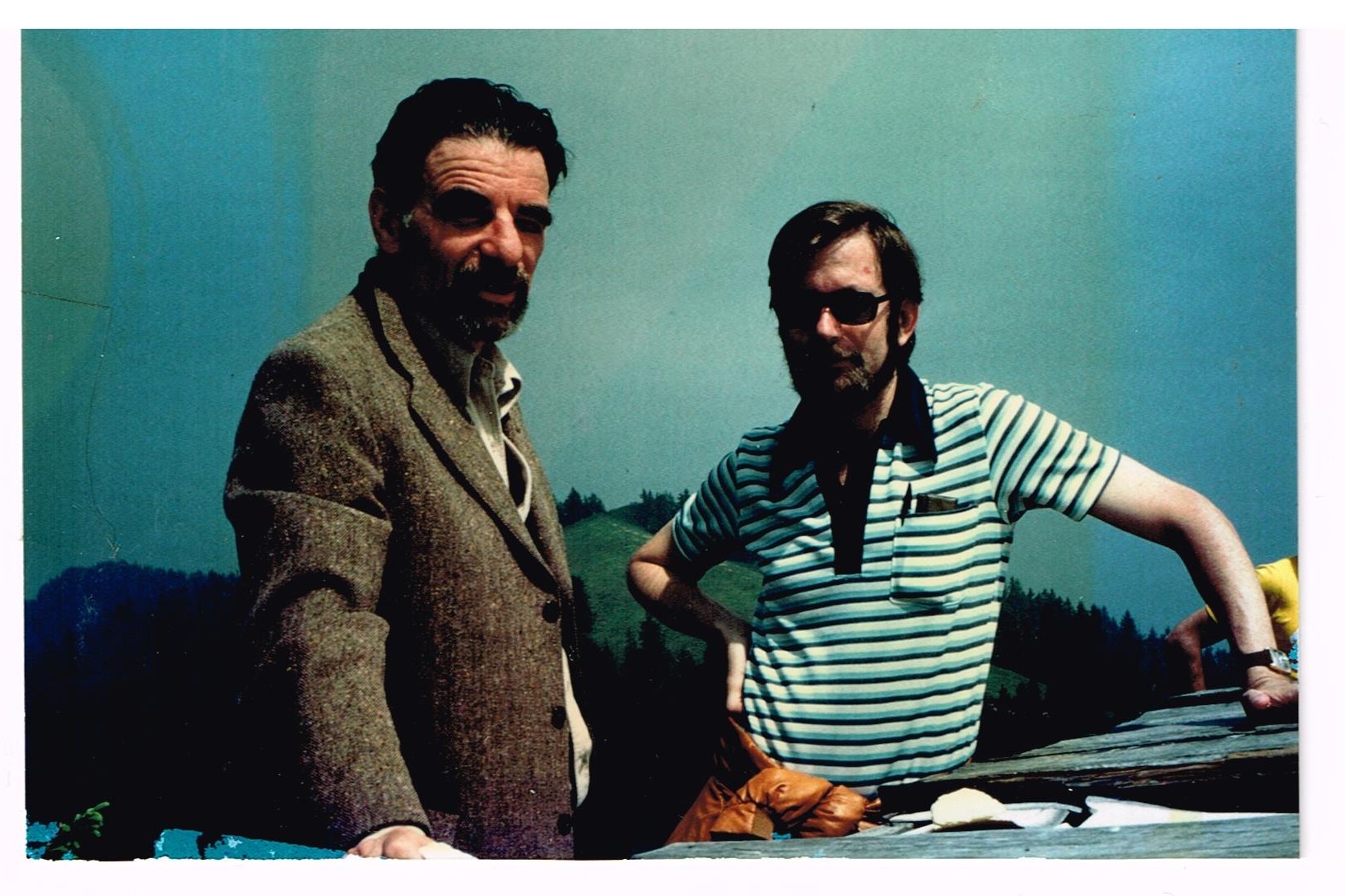}
\caption{Alex and the author: Photo taken on Alex's camera during a hike in the mountains near Z\"urich, circa 1976.}
\label{mount}
\end{figure}

After Alex's visit to Cornell, I visited him many times at the IBM Z\"{u}rich Laboratory in R\"{u}schlikon, where he usually took me for long hikes in the mountains, during which we discussed physics ad many other topics (see Fig. \ref{mount}). One freezing night he also drove me (with Tony Schneider) to an ice hockey match, in his self maintained nice car. The same car also took us to the 4th European Meeting on ferroelectricity in 1980  on the beach of Portoro\v{z}, Yugoslavia (now Croatia), where we both presented talks~\cite{EMF}.
  Alex's 60th birthday was celebrated in a  conference in Z\"{u}rich,  in which my talk's title was ``My life with Alex M\"{u}ller and the perovskites". Unfortunately, I cannot find the old-fashioned transparencies used in that talk.

  Alex also visited Tel Aviv many times. In particular, during one of his visits in the 1980's he discussed superconductivity in granular aluminium with my colleagues Ora Entin-Wohlman and Guy Deutcher~\cite{GD}. He later `blamed' these discussions as one of the reasons for his research on the high-temperature superconductors. I tried to tell him that his work on the perovskites is great, and that he should continue it, but obviously he knew better (I had a similar experience with Heini Rohrer, of the same IBM laboratory, with whom I collaborated on dilute antiferromagnets before he switched to tunneling electron microscopes and the Nobel prize; work on critical phenomena seems to be a good preparation for this prize...). However, Alex told me about the superconductivity in the cuprates before his paper with Bednorz was published, and this enabled some of us to publish the first magnetic model for pairing and the first phase diagram of the cuprates, including a spin glass phase between the antiferromagnet and the superconductor~\cite{HTC,SG}.

 In 1987 I had dinner with knowledgeable Swedes in Oslo, and I concluded that the Swedish academy will call Alex and Bednorz the next day, to inform them about their Nobel prize. I immediately called Alex, told him to stay near the phone the next day, and got his permission to nominate him as an honorary doctor of Tel Aviv University. Indeed, he received this honorary degree in 1988, joining 10 other honorary degrees all over the world (his official letterhead said ``Prof. Dr. Dr. h. c. mult. K. A. M\"{u}ller."). When I told him that this degree entitles him to participate in the board of governors of the university, he took this seriously, ad even expressed an opinion on who should be elected as the university's next president. In 1988 I also attended the ceremony in which Boston University gave him the same honor.

\begin{figure}
\centering
\includegraphics[width=0.4\textwidth]{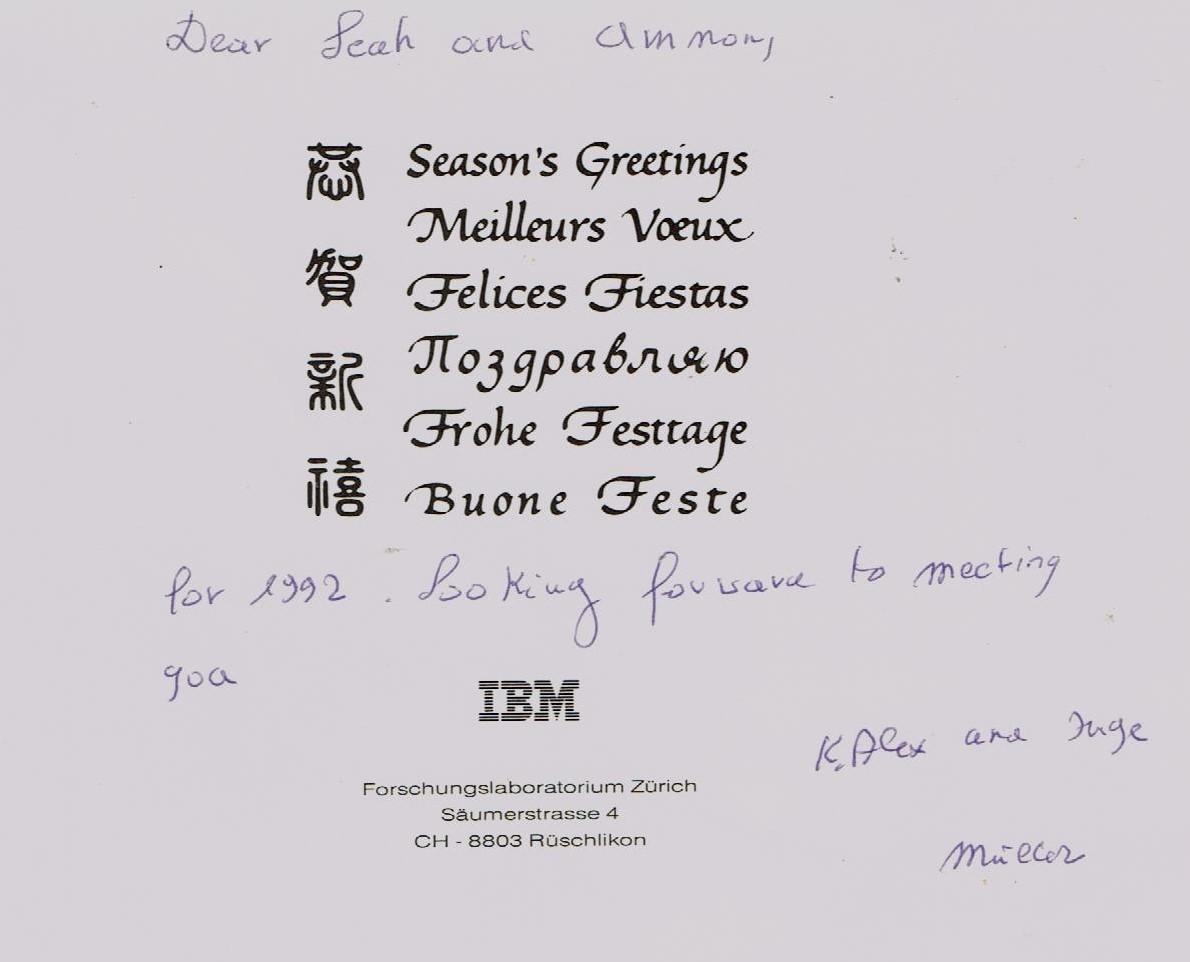}
\caption{A new year card from Alex in 1992, just before visiting Tel Aviv with his daughter Sylvia.}
\label{1992}
\end{figure}

Every year my wife and I got a new year card from him and from his wife Inge, Fig. \ref{1992}. I met Inge  many times in their home in Hedigen and in Tel Aviv. After his retirement, Alex shifted interests to philosophy ad psychology (as before, maybe he knew better?), and our communications became more rare. Sadly, I missed the chance to communicate to him our very recent results on the puzzles of the perovskites, which are described at the end of this paper. I like to think that he would enjoy seeing that - even fifty years after our first meeting - his discussions with me are still alive and fruitful.

\subsection{II. The phase transitions in the perovskites }
\label{puz}

\begin{figure}
\centering
\includegraphics[width=0.2\textwidth]{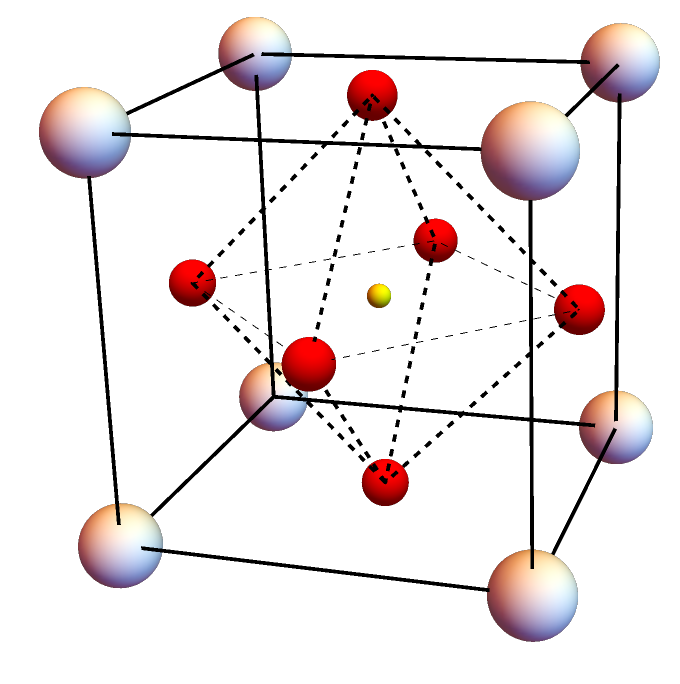}\ \ \ \ \ \includegraphics[width=0.2\textwidth]{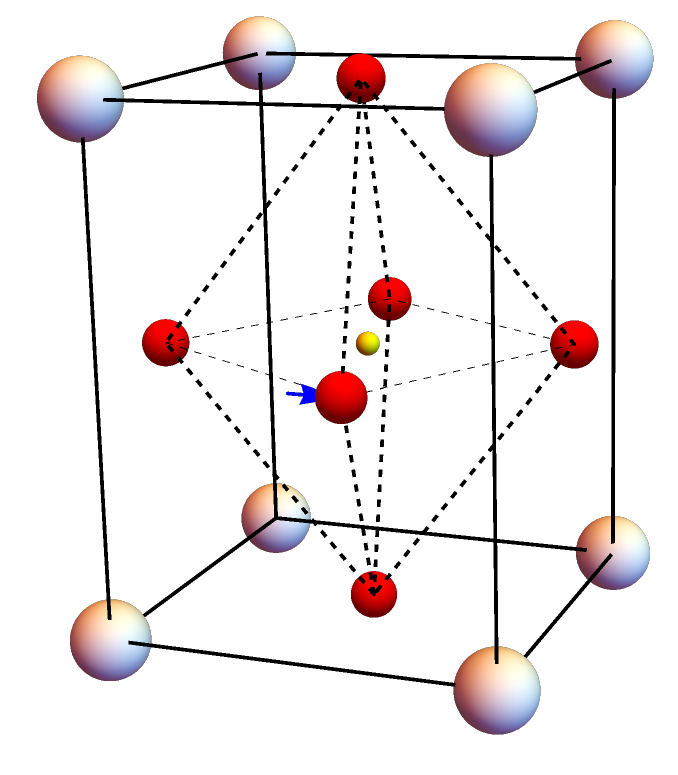}
\caption{(color online) The cubic (left) and tetragonal (right) unit cells in SrTiO$_3$ (the latter shows only half the cell: neighboring cells rotate in opposite directions). Large, intermediate and small spheres correspond to Sr, O and Ti ions, respectively. The dashed lines represent the octahedra, which rotate around the vector ${\bf Q}$, lying along the vertical axis (the O ions in the central horizontal plane move as indicated by the arrow). When ${\bf Q}$ is along a diagonal of the cube, the octahedra rotate around that  diagonal, and the  unit cell  is stretched along ${\bf Q}$, causing a cubic to trigonal transition (as in LaAlO$^{}_3$).}
\label{Fig1}
\end{figure}

Perovskite materials exhibit intriguing physical properties, and have been extensively explored for both practical applications and theoretical modeling~\cite{perov}. In particular, perovskites like SrTiO$^{}_3$ and LaAlO$^{}_3$ play important roles in modern solid state applications~\cite{STO-LAO}. Indeed, these materials continue to be in the center of much currnet research~\cite{revSTO}. At high temperatures,  perovskites usually have a cubic structure (left panel, Fig. \ref{Fig1}).  As the temperature $T$ decreases, some perovskites undergo an antiferrodistortive structural transition from the cubic to a  lower-symmetry structure, via a rotation of the oxygen (or fluorine) octahedra:  SrTiO$^{}_3$, KMnF$^{}_3$, RbCaF$^{}_3$  and  others undergo a cubic to tetragonal transition, see Fig. \ref{Fig1}. In contrast, LaAlO${}_3$, PrAlO$^{}_3$, and  NdAlO$^{}_3$, undergo a cubic to trigonal transition.  As first found by Alex, the  order parameter of these transitions is related to the rotations of the oxygen octahedra in the unit cell~\cite{KAM68}. For the cubic to tetragonal transition, the octahedra rotate around a cubic axis and the order-parameter vector
 ${\bf Q}$ (a.k.a. the rotation vector) is along that axis (with a length proportional to the rotation angle, which is deduced from the EPR spectra). For the cubic to trigonal transition, ${\bf Q}$ is along a cubic diagonal.
 Similar rotations (around a tetragonal axis) occur in  double perovskites, e.g., the tetragonal to orthorhombic transition in the Bednorz-M\"{u}ller parent high-temperature superconductor La$^{}_2$CuO$^{}_4$~~\cite{axe}.  Honoring Alex, Ora Entin-Wohlman and I placed the unit cell of this material on the cover of our solid state book~\cite{book}.

%%%%%%%%%%%%%%%%%%%%%%%%%%%%%%%%%%%%%%%%%%%%%%%%%%%%%%%%%%%%%%%%%%%%%
%%%%%%%%%%%%%%%%%%%%%%%%%%
%%%%%%%%%%%%%%%%%%%%%%%%%%%%%%%%%%%%%%%%%%%

 The  behavior of a system at the vicinity of its transition temperature $T^{}_{c}$ can be expressed by  critical exponents. When the transition at $T^{}_c$  is continuous, the correlation length diverges as $\xi\propto |t|^{-\nu}$ and the order-parameter approaches zero (for temperatures $T<T^{}_c$) as $|\langle{\bf Q}\rangle|\propto |t|^\beta$, where $t=T/T^{}_c-1$. The critical exponents $\nu$ and $\beta$ are expected to be universal, i.e., having the same values for many physical systems which share the same symmetry. The exponents describing other physical properties, e.g., $\alpha$ and $\gamma$ for the specific heat and for the order parameter susceptibility, are obtained via scaling relations, $d\nu=2-\alpha=2\beta+\gamma$, where $d$ is the dimensionality.

 As described in the next section, the experiments of Alex and others raised several puzzles:

 (1) What is the critical behavior of the cubic to tetragonal and of the cubic to trigonal phase transitions? One had to distinguish between three possibilities: (a) Both belong to the same universality class, of the isotropic Heisenberg model. (b) One (or both) belong to the universality class associated with the cubic fixed point of the renormalization group. (c) One (or both) turn first order.

 (2) Some of the transitions from cubic to tetragonal seem to be first order, or to be close to a tricritical point~\cite{1st}. This does not seem to happen for the cubic to trigonal transitions. Why?

 (3) Applying uniaxial stress reduces the number of components of the critical order parameter, from 3 to 1 or from 3 to 2. The 3-component critical point mentioned above then becomes a multicritical point~\cite{BA75}. Is this multicritical point bicritical, tetracritical or triple?

\subsection{III. Theory versus experiments}

{\bf 1958}: Alex developes the EPR measurements in SrTiO$^{}_3$~\cite{phd}.

{\bf 1968}: Alex, Berlinger and Waldner use EPR to measure the rotation angles of the octahedra in both  SrTiO$^{}_3$ and LaAlO$^{}_3$, and identify similarities between their temperature dependences~\cite{KAM68}. All of Alex's EPR experiments were done together with his long-time technician, Walter Berlinger.

{\bf 1970}: Alex, Berlinger and Slonczewski measure the structural transition in SrTiO$^{}_3$ under [111] uniaxial stress, and observe a cubic to trigonal phase transition (in contrast to the cubic to tetragonal transition at zero stress)~\cite{KAM70}. Following Ref. \cite{ST}, the Landau theory free energy for this transition is written as
\begin{align}
\widetilde{U}&=\frac{1}{2}K(T)Q^2+A'Q^4+A'^{}_n\sum_{i<j}Q_i^2Q_j^2\nonumber\\
&-b^{}_e\sum_i T^{}_{ii}(3Q_i^2-Q^2)-b^{}_t\sum_{i<j}T^{}_{ij}Q^{}_iQ^{}_j,
\label{1}
\end{align}
where $Q^2=|{\bf Q}|^2=\sum_{i=1}^3 Q_i^2$, $K(T)\propto(T-T_c^{(0)})$, with $T_c^{(0)}$ being the transition temperature at  zero stress, $A'$ reflects the cubic symmetry quartic term, $T^{}_{ij}$ are the components of the stress tensor and the coefficients $b^{}_e$ ad $b^{}_t$ represent the couplings of the stress tensor to the order parameter components $Q^{}_i$.
The stress $p$ along [111], for which $T^{}_{ij}\equiv -p/3$ for all $i,j$, `prefers' ordering of ${\bf Q}$ along [111], hence the trigonal structure. In this Landau theory, the transition temperature increases linearly with the stress, $T^{}_c(p)-T_c^{(0)}\propto p$.

{\bf 1971}: Alex and Berlinger fit the data for the rotation angles of the oxygen octahedra in both SrTiO$^{}_3$ and LaAlO$^{}_3$ to a crossover from the mean-field exponent $\beta^{}_{MF}=1/2$ (far from $T^{}_c$) to the critical $|{\bf Q}|\sim|t|^\beta$, with $\beta=0.33\pm 0.02$~\cite{KAM71}.

\begin{figure}[htb]
\vspace{-1.4cm}
\includegraphics[width=0.56\textwidth]{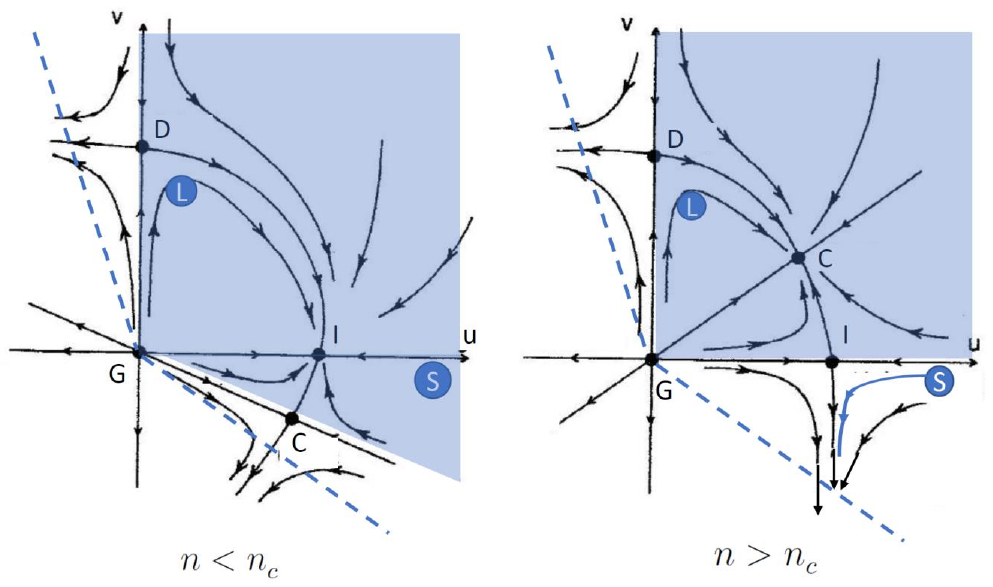}
\vspace{-8.7cm}
\caption{(color online)  Schematic flow diagram and FPs for the cubic model,  adapted from Ref. \onlinecite{DG}. %Left: $n<n^{}_c$. Right: $n<n^{}_c$.
$G=$Gaussian, $I=$isotropic, $D=$Decoupled (Ising) and $C=$Cubic FP's. S=initial point for SrTiO$^{}_3$. L=initial point for LaAlO$^{}_3$ (these locations were estimated by Alex {\it et al.}~\cite{KAM-NATO}). The dashed lines represent the stability edges, $u+v=0$ (for $v<0$) and $u+v/n=0$ (for $v>0$), below which  the free energy in Eq. (2) is stabilized by the terms of order $|{\bf Q}|^6$, and the transitions are first-order. The shaded areas are the regions of attraction of the stable FP's ($I$ on left and $C$ on right). }
\label{fp}
\end{figure}

{\bf 1973}: The author (AA) uses the renormalization group to analyze the critical behavior of cubic systems in $d=4-\epsilon$ dimensions~\cite{AA1973}. Basically, one turns the Landau free energy (\ref{1}) into a Ginzburg-Landau-Wilson (GLW) free energy density, by adding  a gradient term, $|{\boldmath{\nabla}}{\bf Q}({\bf r})|^2/2$,
whose coefficient is normalized to $1$ (and kept equal to $1$ under renormalization). This term represents the interactions, capturing the long range correlations~\cite{WF,WK,DG,MEF}. In the analysis, the terms with $A'$ and $A'^{}_n$ in Eq. (\ref{1}) were replaced by $uQ^4+v\sum_i Q_i^4$.
This analysis yielded four competing fixed points: Gaussian (G), Decoupled Ising (D), Isotropic Heisenberg (I) and Cubic (C), see Fig. \ref{fp} (adapted from Ref. \cite{DG}).
This figure has been  reproduced by  other authors, e.g., Refs. \onlinecite{6loops,eps6,vic-rev}, and included in textbooks~\cite{CL}.
As the figure shows, the relative stability of the fixed points depends on a dimensionality dependent number of order parameter components, $n^{}_c(d)$.
Up to third order in $\epsilon$, it was then concluded that $n^{}_c(3)>3$, and therefore the stable fixed point is the isotropic one. Assuming the initial values of the parameters, S and L, both materials are within the (shaded) region of attraction of the isotropic fixed point (on the left hand side of the figure), and therefore  both the cubic to tetragonal and cubic to trigonal phase transitions should be continuous, exhibiting the Heisenberg critical exponent $\beta^{}_I\simeq 0.377$. Although this scenario supports the apparent universality seen by Alex {\it et al.} in 1971, the measured common exponent $\beta$ disagrees with this isotropic value.

{\bf 1974}: Alex visits Cornell, and discusses with AA and ADB the discrepancy between the theoretical and experimental exponents. Noting that the experiments were done on single crystals which were polished along one axial direction, they concluded that the polishing introduced internal stresses, like $T^{}_{11}$ in Eq. (\ref{1}). This implied a crossover from the isotropic critical behavior to that of the Ising model, with only one critical component of the order parameter.

\begin{widetext}
\begin{figure*}[htb]
\vspace{-1cm}
\includegraphics[width=.8\textwidth]{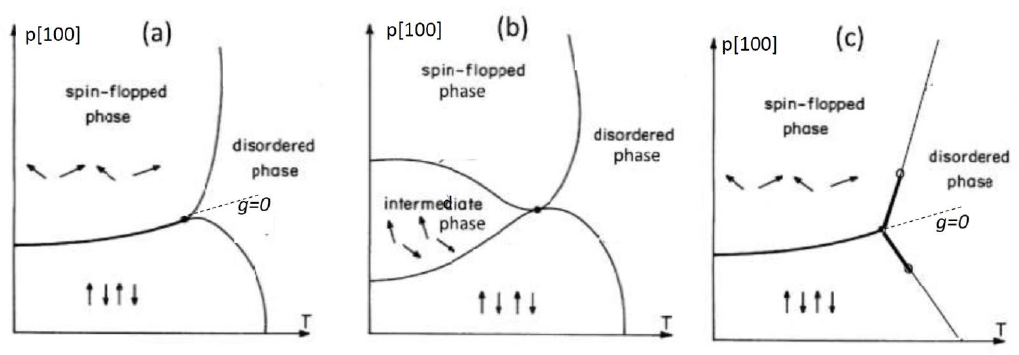}
\vspace{-15.3cm}
\caption{  Possible phase-diagrams for the perovskites under [100] uniaxial stress. (a) Bicritical phase diagram.  (b) Tetracritical phase diagram. (c) Diagram with a triple point (see text).  Thick lines - first-order transitions. Thin lines - second-order transitions. The first-order transition lines between the ordered phases and the disordered paramagnetic phase end at tricritical points (small empty circles). After Refs. \onlinecite{BA75,mukamel}.  }
\label{fd}
\end{figure*}
\end{widetext}

{\bf 1974}: AA and ADB publish a theoretical paper, presenting the bicritical phase diagram expected for SrTiO$^{}_3$ under (positive and negative) uniaxial stress. The 1971 fits by Alex and Berlinger are conjectured to result from a crossover between the isotropic and the Ising behavior~\cite{AB1}. This implied that the polished sample experiments were done at a point on the lower critical line in Fig. \ref{fd}(a).

{\bf 1975}: Alex and Berlinger apply uniaxial stress along [100] and [110] (the latter is equivalent to negative stress along [100]), and confirm the above AA+ADB conjectures (including the first measurement of the bicritical phase diagram for uniaxially stressed SrTiO$^{}_3$). They also fit the 1971 data to $\phi=\phi^{}_0|t|^{\beta(1)}[1+b^{}_1|t|^x]$, with the Ising exponent $\beta(1)=0.315$ and with the correction exponent $x=0.50\pm 0.05$~\cite{KAM2}. The correction term was needed, because the data were not taken very close to $T^{}_c$ and were in the middle of the crossover from the bicritical point to the Ising behavior. They also showed that the shift of $T^{}_c$ due to the stress is not straight, as in the Landau theory, but rather described by the crossover exponent $\phi\simeq 1.25$, as predicted by the renormalization group.

{\bf 1975}: ADB and AA publish a detailed analysis of the bicritical versus the tetracritical phase diagrams for perovskites under stress, based on the renormalized stress-dependent terms in Eq. (\ref{1}) ~\cite{BA75}. They also show that the multicritical point should be bicritical [with a first order `flop' transition between the two ordered phases, Fig. \ref{fd}(a)] for $v<0$, i.e., for the cubic to tetragonal transition under [100] stress, but it should turn tetracritical [with an intermediate `mixed' phase between the above phases, Fig. \ref{fd}(b)] for $v>0$, i.e., for the cubic to trigonal transition under [100] stress.

{\bf 1976}: AA shows that when $v\ne 0$ (either at the cubic fixed point or for effective exponets, which are measured during the renormalization group flow), the crossover exponent due to the uniaxial pressure has different values for stress along [100] or along [111]~\cite{PL76}. These predictions have not yet been tested experimentally (but see below).

{\bf 1977}: AA visits IBM Z\"urich, and joins Alex and Berlinger in a detailed study of SrTiO$^{}_3$ under [111] stress~\cite{AAKAM}, continuing the 1970 work by Alex, Berlinger and Slonczewski~\cite{KAM70}. Below the transition from cubic to trigonal, one encounters a first order transition in which the order parameter components in the (111) plane order. Following Ref. \cite{domany}, they show that the three-fold rotational symmetry of these order parameters yields a three-state Potts model, and they measured (and compared with theory) the exponent describing the growth of the order parameter discontinuity, as a power of the distance from the multicritical point.

{\bf 1979}: AA and ADB visit IBM Z\"urich. This visit resulted in two back-to-back papers~\cite{Lifshitz,KAMLifshitz}. Several perovskites undergoing a cubic to tetragonal transitions had been shown to undergo a first order transition, or a vicinity to a tricritical points (see references in \cite{Lifshitz,1st}).
The theoretical paper interpreted this fact as coming from the strong cubic anisotropy of the interactions: due to the weak coupling between octaherda along the axis of their tetrahedral (alternating) rotation, the coefficient of the quadratic terms in the GLW free energy density in Fourier space, $U^{}_{2,i}({\bf q})Q^{}_i({\bf q})Q^{}_i(-{\bf q})$, have the flat Lifshitz form, $U^{}_{2,i}({\bf q})=r^{}_i+q^2_{\perp,i}+aq^{2L}_i$ (with $q_{i,\perp}^2=q^2-q_i^2$). This yields a critical and a tricritical behaviors with the exponents of the uniaxial Lifshitz exponents. The experimental paper~\cite{KAMLifshitz} confirmed these predictions for uniaxially stressed RbCaF$^{}_3$. Note that such cubic terms in the dispersion relation, $U^{}_{2,i}=r+q^2+f^{}_0(q^{}_i)^2$, were already discussed in Refs.~\cite{AAdip,natt}. Although $f^{}_0$ is weakly irrelevant near the isotropic value $f^{}_0=0$, one expects a crossover to the Lifshitz behavior when $f^{}_0\rightarrow -1$. Increasing $-f^{}_0$ pushes $v$ to more negative values, towards the stability line.

{\bf 1981-3}: AA and Daniel Blankschtein find more options for tricritical points on the bicritical phase diagram, resulting from sixth order terms like $Q^6$~\cite{DB}. Experimental realizations of these predictions for the perovskites were reviewed by Fossheim \cite{Fossheim}.

{\bf 1983}: Alex and collaborators~\cite{KAM-NATO} measured the tetracritical phase diagram of LaAlO$^{}_3$ under [100] stress, confirming the predictions of \cite{BA75}. Also, based on the mean-field region of the experiments, they estimated the `initial' Landau parameters  to be $\{u,v\}^{}_{\rm S}\cong \{1.91, -0.068\}$ for SrTiO$^{}_3$ and  $\{u,v\}^{}_{\rm L}\cong 0.06\pm 0.06,~ 0.68\pm 0.06\}$ for LaAlO$^{}_3$, all in cgs units divided by $10^{43}$. These rough values (which should be improved!) are shown in Fig. \ref{fp} by S and L. As discussed below, they fit beautifully with our expectations: SrTiO${}_3$ has a small initial negative $v$, and we predict that it will  flow  quickly parallel to the horizontal axis towards the universal asymptotic line and the isotropic fixed point before turning downwards.

\subsection{IV. Developments after 2000}

The value of $n^{}_c(3)$ has been a topic of very much activity, ever since \cite{AA1973}. Some of the results up to 2002 were summarized in the review paper \cite{vic-rev}. Practically all the calculational methods yield $n^{}_c(3)\cong 2.9\pm0.1<3$. Very recent results confirm this conclusion even more accurately, see e.g. \cite{MC,boot}. This has far reaching consequences. In particular,  the correct renormalization group flow diagram is the one on the right hand side of Fig. \ref{fp}. The stable fixed point is the cubic one, and the phase diagram of LaAlO$^{}_3$ under [100] uniaxial stress is the tetracritical one, with cubic exponents.  Similar conclusions follow for the multicritical point with $5=2+3$ components, which was thought to be described by the decoupled fixed point (for that case)~\cite{AAcom}, but see below.

In contrast, SrTiO$^{}_3$ has $v<0$ ( the point S in Fig. \ref{fp}), and therefore it cannot reach the cubic fixed point. Therefore, the cubic to trigoal and the cubic to tetragonal transitions {\it cannot belong to the same universality class!} As discussed below, the apparently similar exponents observed for these two transitions result from the slow renormalization group flows, which leave the point S close to the isotropic fixed point, while L flows to the cubic fixed point - whose exponents are very close to the isotropic ones. These conclusions created a new puzzle, for SrTiO$^{}_3$ under [100] stress~\cite{puzzle}: Practically all the experiments (e.g., \cite{KAM71}) exhibited  bicritical phase diagrams. However, as seen in Fig. \ref{fp}, the point S is at a small $v<0$. Therefore, for $n^{}_c(3)<3$ it {\it cannot} flow to the cubic fixed point. As this schematic flow diagram shows, this point must flow to more negative values of $v$, eventually crossing the mean field stability line
$u+v=0$. The renormalization group iterations can be stopped after $\ell$ iterations, when the renormalized correlation length $\xi(\ell)=\xi(0)/e^\ell$ becomes of the order of the lattice constant. At that point, most of the fluctuations have been eliminated, and one can use the Landau theory, witn the renormalized parameters $u(\ell)$ and $v(\ell)$. If these values are below the stability line, this theory yields a first order transition. By continuity, the two transition lines leaving this multicritical point also become first order, turning second order only at a finite distance from this point~\cite{mukamel}.
The multicritical point thus becomes a triple point, as in Fig. \ref{fd}(c).
However, the experiments on SrTiO$^{}_3$ do {\it not} show a triple point! Instead, they show a bicritical point, with an apparent second order. This puzzle remained open until 2022.
The same issue also arose for many other apparent bicritical points (e.g., \cite{AAcom}).

\subsection{V. The 2022 solution to the puzzle \\ (with Entin-Wohlman ad Kudlis)}

To solve this puzzle, Ora Entin-Wohlman and I developed a new approach, also with Andrey Kudlis (who is a expert on the resummation techniques)~\cite{puzzle,puz1,puz2,puz3}. Most of the accurate calculations, mentioned above, concentrated on calculating the critical exponents, at the fixed points, and not the renormalization group flow diagrams away from these fixed points. For example, the high-order $\epsilon-$expansions do not connverge, and the fixed points and the critical exponents required a resummation of their series~\cite{eps6}. To derive the renormalization group recursion relations (a.k.a. the flow equations in parameter space), one would need to derive long expansions of the flow equations in $u(\ell)$ and $v(\ell)$, and resum them for each pair of these parameters. Instead, we noticed that the cubic fixed point is very close to the isotropic fixed point (for the same reason, $n^{}_c(3)$ is very close to $3$), and therefore we decided to expand the flow equations around the isotropic fixed point. The small deviations from this point allowed us to stop at quadratic order in these deviations, $\delta u(\ell)=u(\ell)-u_I^\ast$ and $v(\ell)$. Group theory shows that the isotropic fixed point has only three independent stability  exponents for the quartic spin terms, and one of them breaks the permutation symmetry in favor of the two ordered phases in the bicritical diagram. Therefore, along the `symmetric line $g=0$, {\it all} the bicritical and tetracritical flow diagrams (including those involving the biconical fixed point instead of the cubic one, for the case of $n=n^{}_1+n^{}_2$ components) require only two independent variables, like $u$ (for the isotropic $Q^4$) and $v$ (for the appropriate combination of the nine quartic terms associated with the spherical harmonics $Y^{}_{4,m}$). The coefficients in the new recursion relations were derived by resummations of the corresponding $\epsilon-$expansions  of derivatives (with respect to $\delta u$ and $v$) at the isotropic fixed point.

At quadratic order, the flow equations can be solved analytically, and the details are included in Refs. \cite{puzzle,puz1,puz2,puz3}. These calculations yielded two main results. First, while the stability exponent for the isotropic $u$ is negative and large, $\lambda^I_u=-0.7967(15)$, that for the symmetry breaking term $v$ is negative but very small, $\lambda^I_v=0.0083(15)$. Therefore, the non-linear variable associated with $\delta u$ decays quickly to zero, as $e^{\lambda^I_u\ell}$, and one is left with a (novel) universal line which describes the flow in the $u-v$ plane. Since $\lambda^I_v$ is so small, the flow away from the isotropic fixed point (in both directions) is very slow, and the point S$[u(\ell),v(\ell)]$ stays close to the isotropic fixed point for many iterations. If the initial correlation length is not very large (i.e., $T$ is not very close to $T^{}_c$), the above point remains above the stability line, and one observes an {\it apparent} bicritical point, with apparent isotropic exponents. However, very accurate experiments, or initial points which happen to be very close to the stability lines, may cross this line, and exhibit the triple point of Fig. \ref{fd}(c). Indeed, KMnF$^{}_3$ and  RbCaF$^{}_3$ do show first order transitions at some finite $t$, implying that they start at larger values of $|v(0)|$. This offers an alternative explanation of their behavior, competing with that of Ref. \cite{Lifshitz}. Only an accurate dedicated experiment may distinguish between these two scenarios. Theoretically, one should repeat our recent analysis including the renormalization group flow of $f^{}_0$.

\begin{figure}[t]
\includegraphics[width=0.6\linewidth]{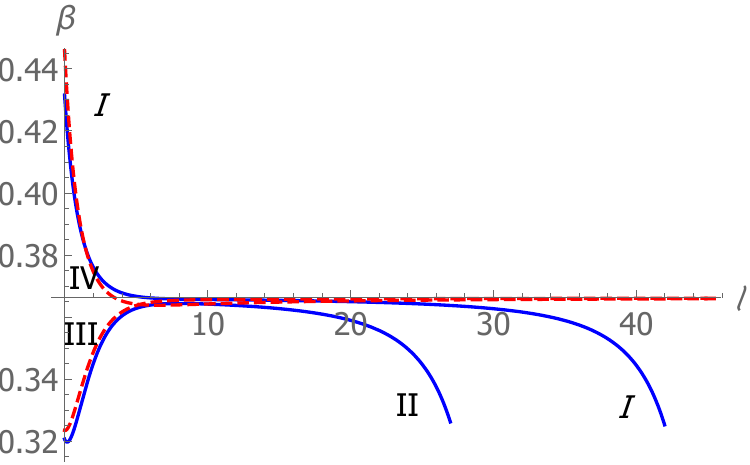}
\caption{%(color online)
 Dependence of the effective critical exponent $\beta$  on the RG flow parameter $\ell$. Different lines correspond to different initial values  $v(0)$, for a fixed $u(0)$.
  In this plot, the trajectories are shown only at $v(\ell)>-.8$, where our quadratic approximation is reasonable. From Ref.~\cite{puzzle}.}
\label{beta}
\end{figure}

Second, once we have the transient $u(\ell)$ and $v(\ell)$, we can put them into the flow equations for the other variables (e.g., the temperature difference $t$ and the uniaxial pressure $p$, and derive the local effective exponents $\beta(\ell)$ and $\phi(\ell)$, with $\ell$ coming from $\xi(\ell)=1$.
Away from the critical point, these exponents remain close to their isotropic values. However, as $\ell$ increases (moving away from the isotropic fixed point), the negative $v(\ell)$ changes quickly towards the stability line, and the effective exponents change significantly. In particular, the effective $\beta(\ell)$ decreases for $v<0$, Fig. \ref{beta}, possibly explaining its low values observed for  KMnF$^{}_3$ and  RbCaF$^{}_3$~\cite{1st}!

\begin{figure}[htb]
\includegraphics[width=0.3\textwidth]{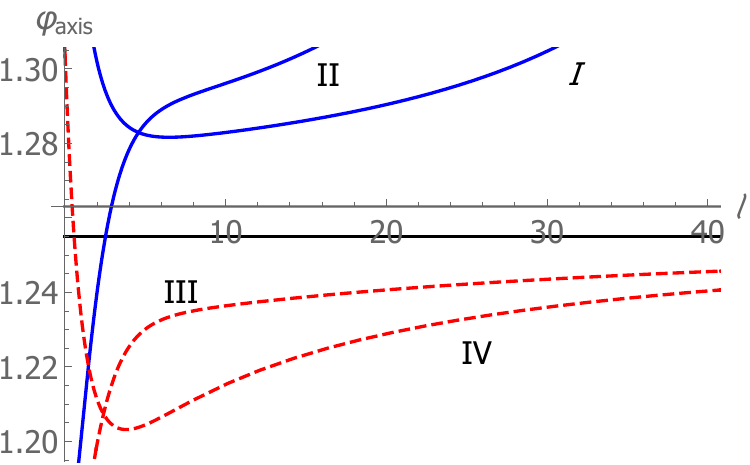}\\
\vspace{5mm}
\includegraphics[width=0.3\textwidth]{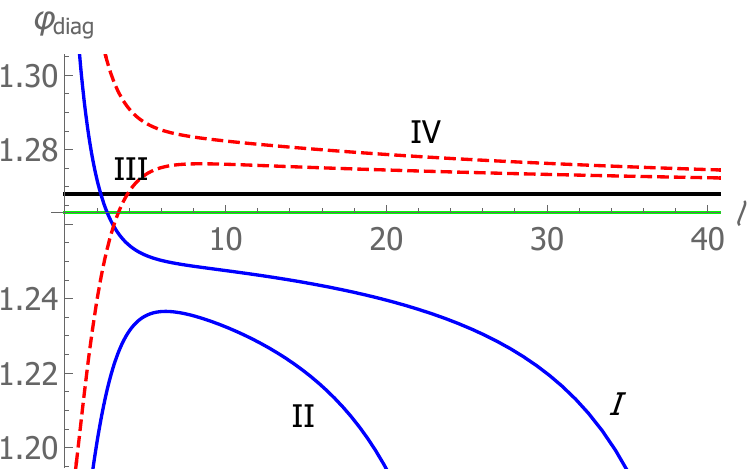}
\caption{(color online)  The effective exponents $\varphi^{}_{diag}(\ell)$ and $\varphi^{}_{axis}(\ell)$ for several initial values of $u$ and $v$, as functions of $\ell$. The horizontal axes (green lines) are at the asymptotic values of the isotropic fixed point, $\varphi^I=1.263$ (which is the same for both panels).  The black horizontal lines show the cubic asymptotic values, $\varphi^C_{axis}=1.255$ and $\varphi^C_{diag}=1.268$. The exponents corresponding to trajectories with $v(0)>0$ (III and IV, dashed lines) approach the asymptotic values of the cubic fixed point, visibly different from the isotropic counterparts. In contrast, those with $v(0)<0$ (I and II, full lines)  initially come close to  these values, but then  turn downward to smaller values, towards the fluctuation-driven first-order transition. From Ref. \cite{puz1}. }
\label{f3}
\end{figure}

In particular, it is interesting to note that the  crossover exponents which characterize the non-straight lines in the multicritical phase diagrams for [100] and [111] pressure \cite{PL76} deviate from their isotropic value in opposite directions, Fig. \ref{f3}. It would be interesting to test these new predictions.

\subsection{V. Conclusions}

Our recent work answers all the questions posed in the beginning:

(1) Contrary to the common belief, that the (cubic) symmetry in the disordered phase should determine the universality class, the cubic to trigonal and the cubic to tetragonal transitions {\it do not} belong to the same universality class. The former belongs to the universality class of the cubic fixed point and the latter should have a fluctuation driven first order transition, with effective exponents which crossover from the mean field values to the isotropic fixed point values, and then change quickly as the system approached this first order transition. Since the cubic fixed point is very close to the isotropic one, intermediate temperature ranges could yield similar effective exponents for both types of transitions, as originally observed by Alex.

(2) The above difference also explains why (only) the cubic to tetragonal transitions approach a first order transition, either if the initial point in the $u-v$ diagram is close to the stability line, or if the initial correlation length is large enough.

(3) Asymptotically (at large initial correlation lengths) the [100] stressed cubic to trigonal phase diagram should be tetracritical, with cubic exponents (that depend o the direction of the uniaxial stress). In contrast, the [100] stressed cubic to tetragonal phase diagram should have a triple point. However, the latter may exhibit an apparent bicritical point, with varying effective exponents, at intermediate temperature ranges.

It should be emphasized that the same conclusions apply to many other phase transitions from cubic phases, e.g., in magnets, ferroelectrics, and more~\cite{puz2}.

As shown, the main difference between the various multicritical phase diagrams concerns the initial sign of the cubic parameter $v$. Alex and I discussed this issue many times, and concluded that a way to vary $v$ experimentally is to use mixed crystals, e.g. Sr$^{}_{1-x}$Ca$^{}_x$TiO$^{}_3$~\cite{mixed}  or a mixture of SrTiO$^{}_3$ with LaAlO$^{}_3$, which is expected to be easy to grow due to their matching lattice constants~\cite{STO-LAO}.  Since  both the isotropic and cubic FP's have $d\nu>2$, randomness is irrelevant~\cite{ABH,rAA} and one expects the same competition predicted above. %\st{The properties of many mixed
It is interesting to note that KMn$^{}_{1-x}$Ca$^{}_x$F$^{}_3$ seems to approach a second-order transition as $x$  increases~\cite{KTCaO}. If the transition is still into the tetragonal structure, this may represent a smaller value of the initial $|v|$  in the dilute case. This may be caused by the larger dimension of the parameter space in the dilute case, which involves many transient iterations until the flow reaches the $u-v$ plane ~\cite{rAA}.

As already said, I am sure that Alex would enjoy these results. I miss our energetic discussions.

\begin{acknowledgments}
I am grateful to Alex M\"uller for his friendship and collaboration. I also thank Alastair Bruce for many old joint papers, and
Ora Entin-Wohlman and Andrey Kudlis for a very fruitful recent collaboration.

\end{acknowledgments}

%%%%%%%%%%%%%%%%%%%%%%%%%%%%%%%%%%%%%%%%%%%%%%%%%%%%%%%%%%%%55
%%%%%%%%%%%%%%%%%%%%%%%%%%%%%%%%%%%%%%%%%%%%%%%%%%%%%%%%%%%%55


\begin{thebibliography}{99}
\bibitem{1st} Examples include, e.g.,  $\beta\sim 1/3,~\sim 0.27,~\sim0.17\pm 0.02$
for  KMnF$^{}_3$~\cite{borsa}, RbCaF$^{}_3$ and NaNbO$^{}_3$~~\cite{NaNbO3}, respectively. Furthermore, some experiments hint that SrTiO$^{}_3$ may be close to a tricritical point ~\cite{garnier,STO1st,Gallardo}, while both RbCaF$^{}_3$ and KMnF$^{}_3$ have  first-order transitions~\cite{Rb,stokka}. Interestingly, some experiments
 exhibit intermediate regions (before the first-order transitions) with effective critical exponents, which are smaller than their isotropic values. This is consistent with our Fig. \ref{beta}.  The  smaller values of $\beta$ in NaNbO$^{}_3$ have also been attributed to the inter-plane weak correlations along the rotation axis~\cite{Lifshitz}. More
 experiments are reviewed e.g. in Refs. \onlinecite{NATO,KAMrev,cowley}.
\end{thebibliography}
\end{document}